\documentclass[twocolumn,showpacs,amsmath,amssymb,superscriptaddress]{revtex4-1}

\usepackage{graphicx} 
\usepackage{dcolumn} 
\usepackage{bm} 

\usepackage{textcomp}
\usepackage{amssymb}
\usepackage{hyperref}

\begin{document}

\title{Performance of Polarization-based Stereoscopy Screens}

\author{Xiaozhu Zhang}
\author{Kristian Hantke}
\affiliation{Max Planck Institute for Dynamics and Self-Organization (MPIDS), 37077 Goettingen, Germany}
\author{Cornelius Fischer}
\affiliation{Georg-August-Universit\"at G\"ottingen,  37077 G\"ottingen, Germany}
\author{Matthias Schr\"oter}
\email{matthias.schroeter@ds.mpg.de}
\affiliation{Max Planck Institute for Dynamics and Self-Organization (MPIDS), 37077 Goettingen, Germany}

\date{\today}

\begin{abstract}
The screen is a key part of stereoscopic display systems using polarization to separate the different channels for each eye.
The system crosstalk, characterizing the imperfection of the screen in terms of preserving the polarization of the incoming signal,
and the scattering rate, characterizing the ability of the screen to deliver the incoming light to the viewers,
determine the image quality of the system. Both values will depend on the viewing angle.
In this work we measure the performance of three silver screens and three rear-projection screens. Additionally, we measure the
surface texture of the screens using white-light interferometry. While part of our optical results can be
explained by the surface roughness, more work is needed to understand the optical properties of the screens
from a microscopic model.
\keywords{stereoscopic projection \and polarization \and screen \and system crosstalk \and ghosting \and surface texture}
\end{abstract}

\maketitle

\section{Introduction}
\label{sec:intro}
Displaying 3D content is not only an important issue in the entertainment industry, it is also of increasing
importance in science where new numeric and experimental methods have created a wealth of three-dimensional datasets.
Many stereoscopic display system are based on polarization filtering:
the visual information for each eye is oppositely polarized, projected to and scattered or transmitted by the screen,
and finally filtered by the viewer's glasses which consist of two polarizers admitting only the correctly
polarized light to each eye \cite{iizuka:06,janssen:08,kim:10}.
There are two options for polarization filtering: linear and circular polarized light. While linear polarizers
are simpler to manufacture, circular polarization has the advantage that head tilting will not impair
the quality of the image

An ideal screen would completely preserve the polarization of the incoming light. However, in practice
there is always some amount of \textquotedblleft ghosting\textquotedblright \, resulting from the change of
polarization at the screen. A measure for ghosting is the system crosstalk $C$.
It is defined as the ratio between the intensity of light that leaks from the unintended channel to the intended one
and the intensity of the intended channel \cite{woods:11}.
According to measurements of Huang {\it et al}. \cite{huang:03}, the maximal acceptable system crosstalk
for a typical viewer to still experience a stereo sensation is 0.1. (Lower values down to $10^{-4}$ can still be
detected by careful visual inspection). While it is also known on a theoretical basis
that the viewing angle will influence the amount of system crosstalk \cite{richards:10}, to our knowledge no
measurements of the angle-dependent system crosstalk of different screen types have been published up to now.
Neither has the question been studied how the inclination angle (between the incoming light from the projector and the
surface normal of the screen) influences the system crosstalk.

A second measure for the quality of a screen is the brightness of the image,
which depends on the amount and angular distribution of the reflectance (for silver screens)
or transparency (for rear-projection screens) of the screen.
For silver screens this is typically quoted as the screen gain, the intensity measured at normal incidence
normalized by the intensity of a Lambertian source \cite{brennesholtz:08}.
Here we measure the angle dependent scattering rate $S$ for both silver screens and rear-projection screens.
$S$ is defined as the ratio of the intensity received by a viewer in a certain angle to the intensity of the incoming
light, normalized by the solid angle.

In this paper we present measurements of the angular dependence of system crosstalk and scattering rate for
three samples of silver screens (labeled SS1 to SS3) and three rear-projection screens (RP1 to RP3).
Additionally, we determine the surface texture of the samples using white-light interferometry; this information
provides some qualitative insight into our optical results.

\section{Experimental Setup}
\label{sec:exp}
\newcommand{\imsize}{0.58\columnwidth}
\begin{figure*}[!ht]
\begin{center}
\begin{tabular}{c   c   c}
{\resizebox{\imsize}{!}{\includegraphics{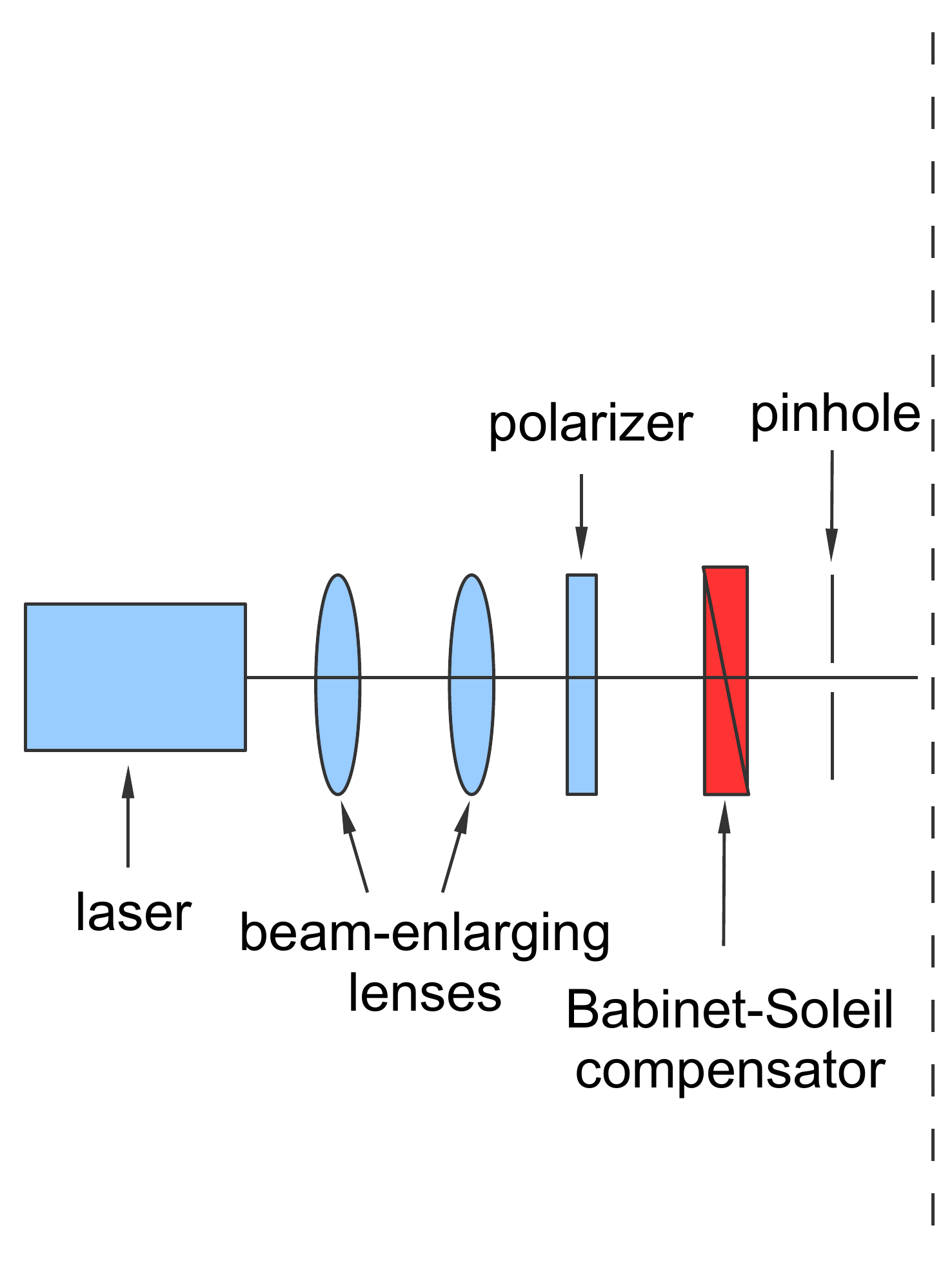}}} \hspace{2cm}&
{\resizebox{\imsize}{!}{\includegraphics{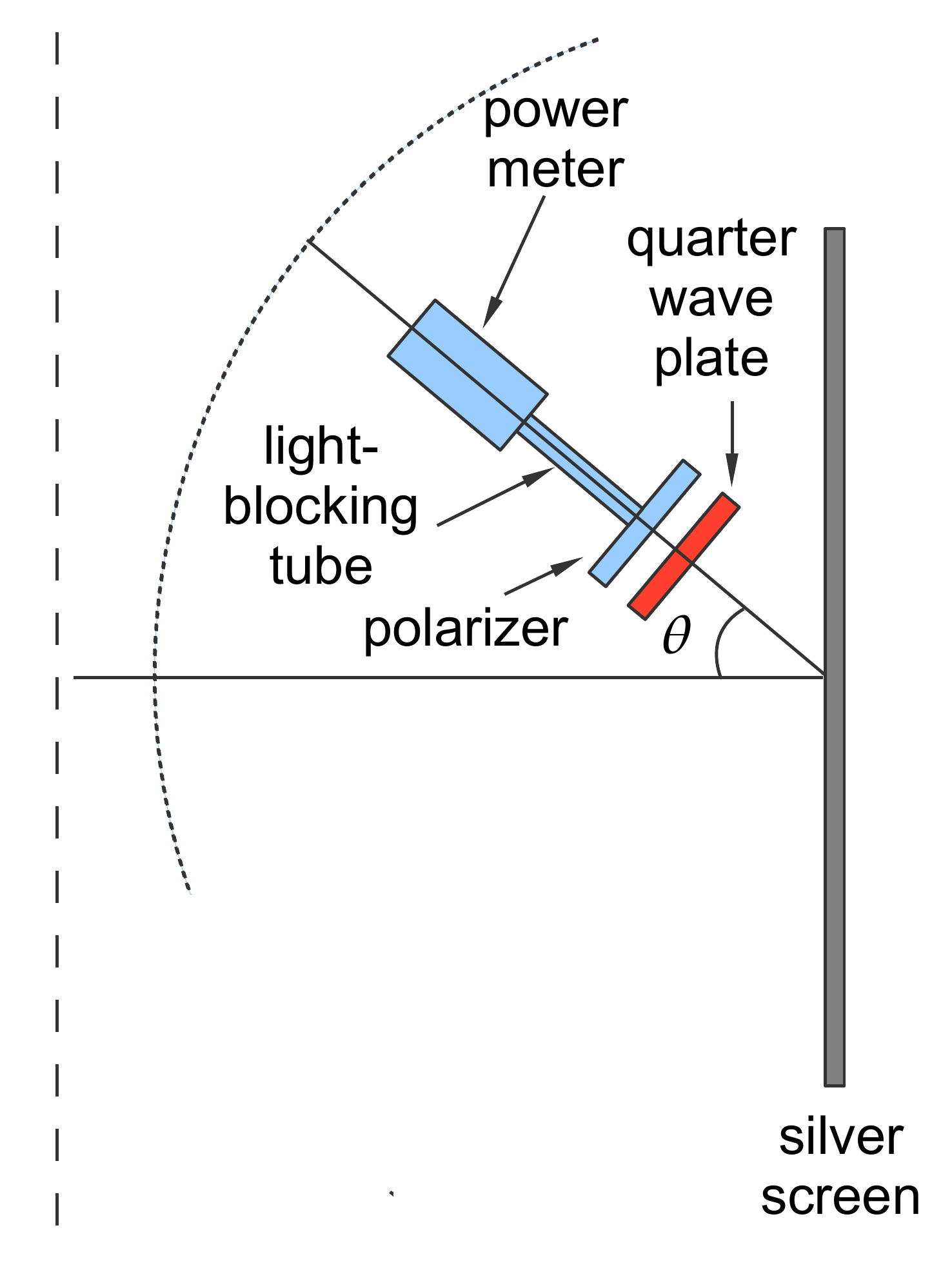}}} \hspace{2cm}&
{\resizebox{\imsize}{!}{\includegraphics{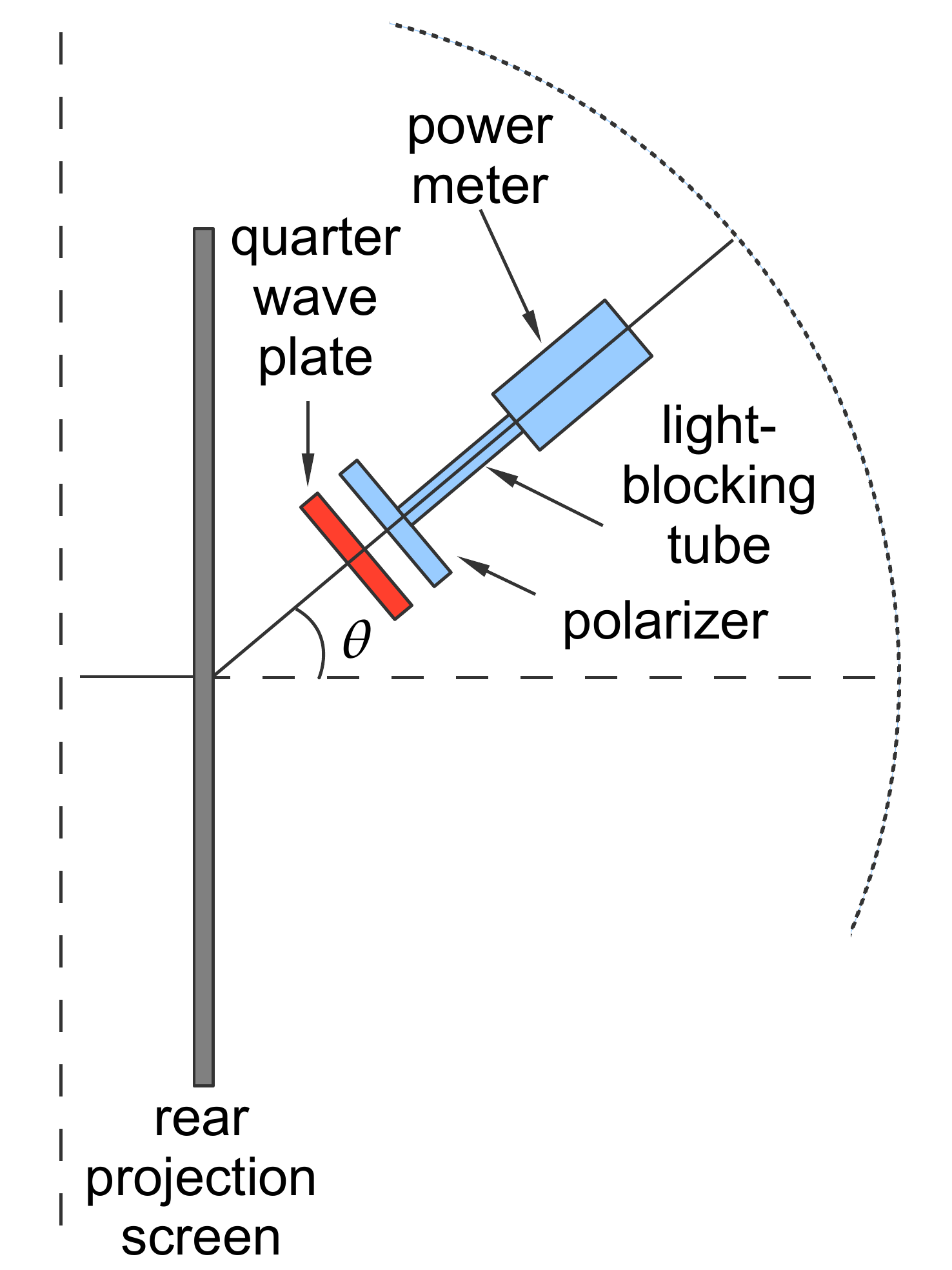}}}\\
(a) & (b) & (c)\\
\end{tabular}
\end{center}
\vspace*{-0.5cm}
\caption{Sketch of the experimental setup using circular polarized light: panels a) and b) are used for silver screens
and a) and c) for rear-projection screens. The generation and the detection of linearly polarized light is
achieved by removing the devices colored in red.}
\label{fig:sketch}
\end{figure*}

Figure \ref{fig:sketch} shows the experimental setup used for measuring the angular dependence of the system crosstalk and
the scattering rate. Diode pumped solid state lasers (DPGL-2050 from Photop and  Verdi V5 SF from Coherent)
with a wavelength of 532nm were used as  light sources for the experiments.
Passing a beam expander, the diameter of the laser beam was increased to 3.4mm (FWHM), whereas the typical size of
structural inhomogeneities on the screen surface is at most a few hundred micrometers as shown below.
This ensured that the measured data for different spots on the screen are reproducible within $\pm 5$\%.

The laser light was linearly polarized by passing a polarizer
or circularly polarized by passing an additional Babinet-Soleil compensator (from B. Halle). The screen sample is irradiated by
the laser at normal incidence and the scattered laser light of the silver and rear projection screens is detected by an detection
unit in reflection (\ref{fig:sketch}b) and in transmission (\ref{fig:sketch}c), respectively. The detection unit consists of a power
meter (PM100D with sensor S130C from Thorlabs), an analyzer and in case of the circular polarization an additional quarter-wave plate
(both from B. Halle). A long and narrow tube was placed in front of the power meter, ensuring that only the photons are detected
that scatter from the irradiated spot on the screen along the viewing axis of the sensor in a solid angle $\Omega$ of 2$\cdot$10$^{-4}$\,sr.
The detection unit was placed on a rotatable rail with the rotational axis being fixed in such a way that the normal viewing axis
of the sensor intercepts always with the illuminated area on the screen during rotation. The viewing angle $\theta$ can be varied
 from -20\textdegree \, to 80\textdegree. For silver screens the range of $\pm$6\textdegree \, is inaccessible in order to not block
the incoming beam.

In both cases of linear and circular polarization the incoming laser intensity $I_{in}$ was measured just in front of the sample.
Furthermore, the intensity of the scattered light $I_{out}$ was measured for the intended channel with the polarization being the
same direction as the incoming one ($I_{out}^{sp}$, analyzer and polarizer parallel) and for the unintended channel with the polarization
being the opposite direction ($I_{out}^{op}$, analyzer and polarizer perpendicular) for different viewing angles $\theta$. From this
data one can compute the crosstalk $C(\theta)$:
\begin{equation}
  C(\theta) = \frac{I_{out}^{op}(\theta)}{I_{out}^{sp}(\theta)}
  \label{eq:C}
\end{equation}
and the scattering rate $S(\theta)$:
\begin{equation}
  S(\theta) = \frac{I_{out}^{sp}(\theta)+I_{out}^{op}(\theta)}{I_{in} \; \Omega}
  \label{eq:S}
\end{equation}

The precision of the measurement for the crosstalk depends strongly on the purity of the initial laser polarization, whereas the
scattering rate is not affected within our measurement precision.
Analyzing the crosstalk without any screen sample (i.e. putting the laser directly in front of the analyzer system)
we found the lower resolution limit in the linear case  to be less than $3 \times 10^{-3}$.
In the circular case the degree of polarization results in a lower resolution limit of
$7.5 \times 10^{-3}$.

The screen samples were obtained from the company Screenlab (Elmshorn, Germany),
their specifications and brand names are listed in table \ref{tab:roughness}.
\begin{table}
\caption{Sample labels, brand names, manufacturer information on gain and transmission, and surface
properties measured by white light interferometry: the root mean square roughness $R_q$ and the ratio
between the surface area and the projected area $F$.}
\label{tab:roughness}
\begin{tabular}{l| c c c| c c}
\hline\noalign{\smallskip}
Sample & brand name  & gain  & transmission  & $R_q [\mu m]$ & $F$  \\
\noalign{\smallskip}\hline\noalign{\smallskip}
RP1 & BS XRP3  &   & 41.8  & 3.8 & 1.03\\
RP2 & WS XRP3  &   & 88.8  &  3  & 1.13\\
RP3 & BS RP2   &   & 41.2  & 4.2 & 1.2\\
\noalign{\smallskip}\hline\noalign{\smallskip}
SS1 & SH120    & 2.4  &  & 5   & 2.3\\
SS2 & SF120    & 2.4  &  & 8   & 2.6\\
SS3 & WA160    & 1.3  &  & 22  & 2.2\\
\noalign{\smallskip}\hline
\end{tabular}
\end{table}

\begin{figure*}[!ht]
\centering
  \includegraphics[width=1\textwidth]{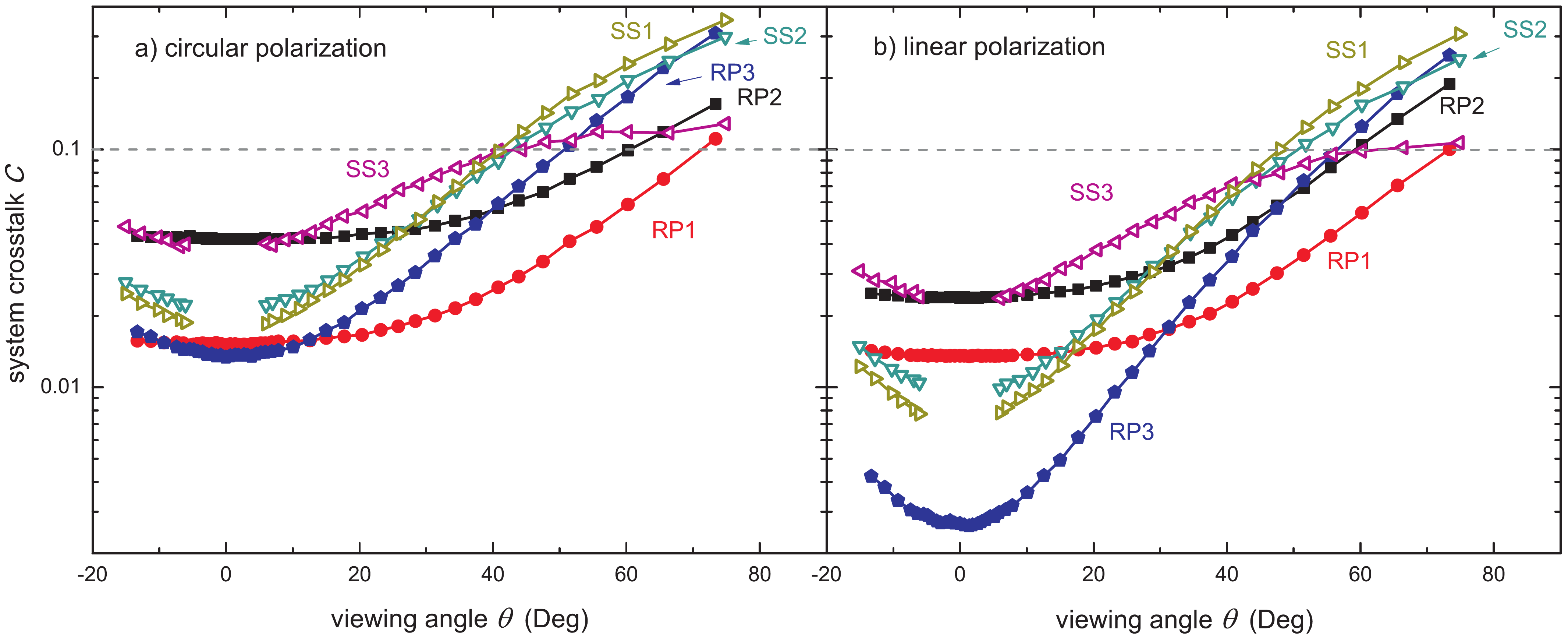}
  \vspace*{-0.5cm}
\caption{Angular dependence of the crosstalk caused by the screen using a) circular polarized light and b) linear polarization.
The grey dash line corresponds to the threshold for still acceptable stereo fusion according to
Huang {\it et al.} \cite{huang:03}. }
\label{fig:crstk}
\end{figure*}

The surface topography of the screens was measured using a ZeMapper whitelight
vertical scanning interferometer (Zemetrics, Tucson, USA):
the focal plane of an interference pattern is vertically scanned through the sample topography,
then a height map is calculated from the collected amplitude maps of the interference patterns.
The vertical resolution of the instrument is better than 1 nm;
the maximum field of view applied in this study is 1.4 mm.
For more information on the instrument see \cite{darbha:10}.
Prior to the measurement the rear projection screens where sputter coated with a
40 nm  gold layer to increase surface reflectivity.

\section{Results}
\label{sec:results}

Figure \ref{fig:crstk} displays the system crosstalk of the six screen samples, both with circular and linear polarized light.
Based on the criterion found by Huang {\it et al.} \cite{huang:03}, all screens allow stereo vision for viewing angles $\theta$ smaller
than 40\textdegree \,  (circular) or 48\textdegree \,  (linear). In practice this range will be smaller due to the additional crosstalk
originating from the glasses and the inclination angle of the incoming light; the latter effect will be described below.

In general, the screens seem to fall into two categories; they are either optimized for a large
range of acceptable crosstalk or a minimized crosstalk at small $\theta$. In both categories
the silver screens are outperformed by the rear projection screens:
RP3 has a smaller $C$ at small $\theta$ than SS1 while RP1 has a broader range of acceptable viewing angles than SS3.

\begin{figure}
\centering
  \vspace*{-0.6cm}
\begin{minipage}{\columnwidth}
\centering
  \hspace*{-1cm}
  \includegraphics[width=1.2\textwidth]{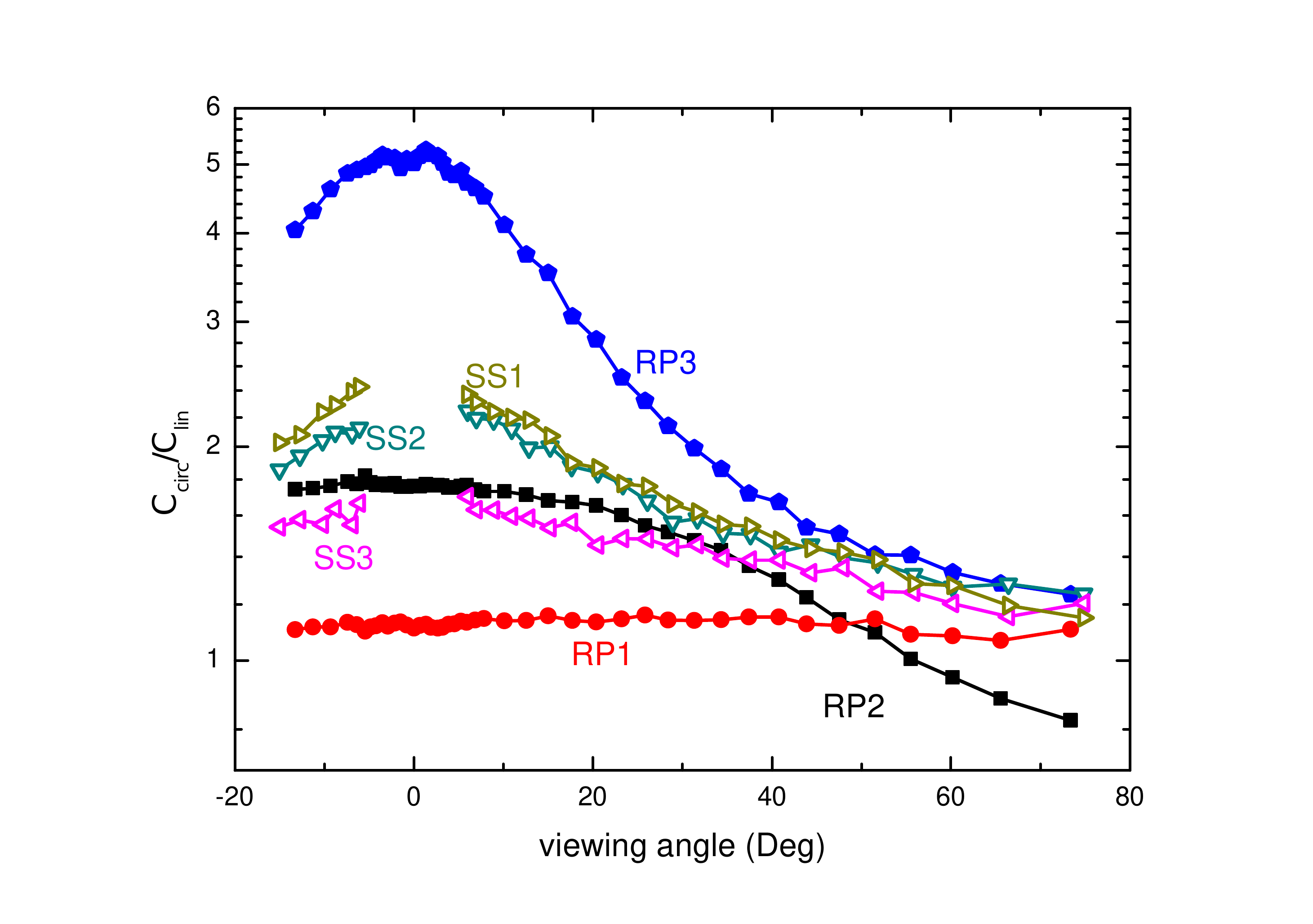}
  \vspace*{-1cm}
\caption{Ratio of crosstalk measured with circular and linear polarization of the incoming light.}
\label{fig:c_ratio}
\end{minipage}
\begin{minipage}{\columnwidth}
  \vspace*{-0.705cm}
  \hspace*{-1cm}
  \includegraphics[width=1.2\textwidth]{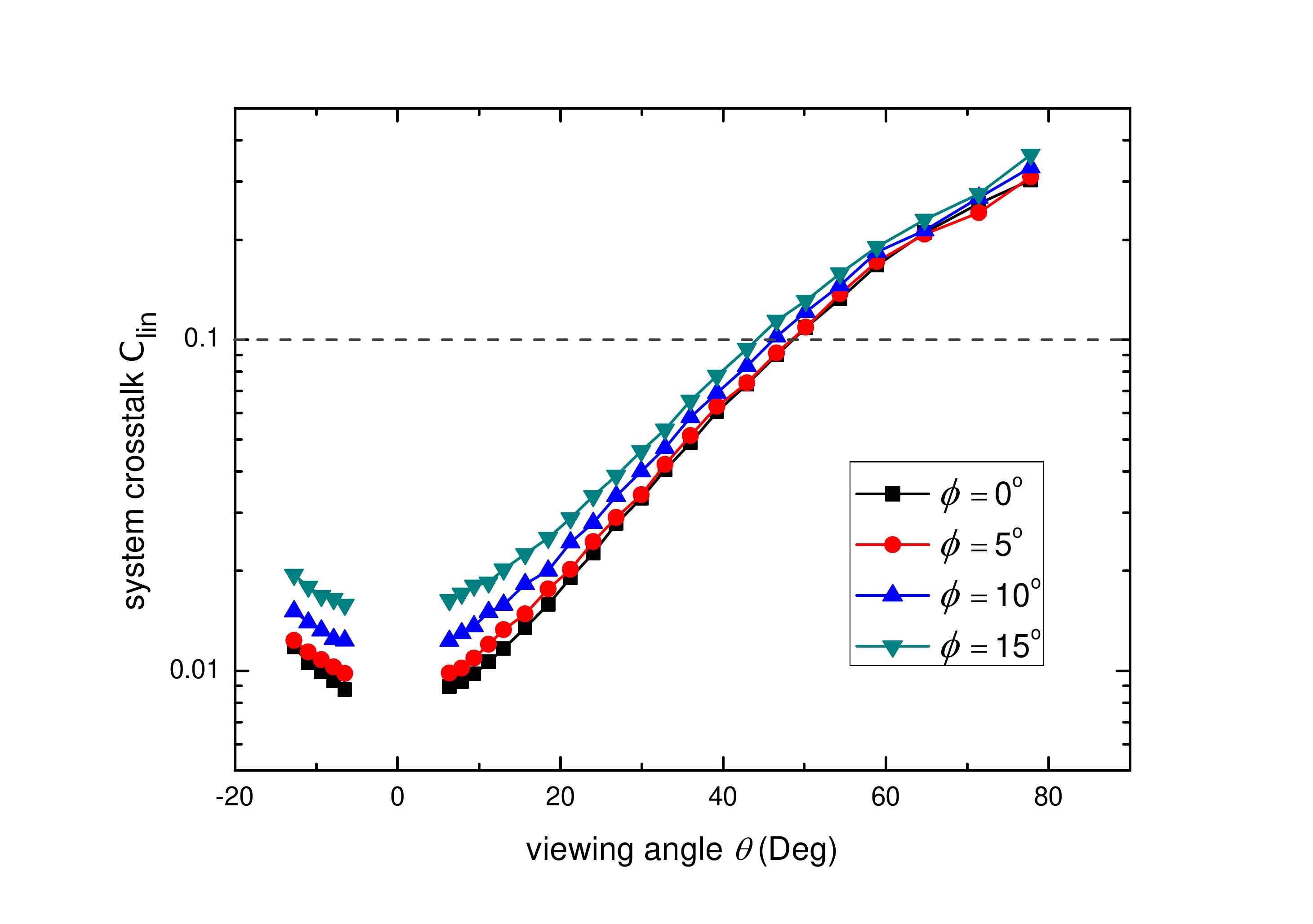}
  \vspace*{-1cm}
\caption{Dependence of the crosstalk on the inclination angle $\phi$.
Measured on screen SS1 using linear polarized light.
}
\label{fig:crstk_incl}
\end{minipage}
\end{figure}

Regarding the polarization mode, linear polarization has for each screen a clear advantage over circular.
$C_{circ}/ C_{lin}$ measured at $\theta$ = 10\textdegree \, varies between 1.1 (RP1) and 4 (RP3)
as shown in figure  \ref{fig:c_ratio}.
Please observe, that our measurements of the crosstalk of RP3 at small angles might be limited by
our experimental resolution.

Under real world conditions it is quite likely that the incoming light itself will have
an inclination angle $\phi$ to the surface normal of the screen. To quantify the additional crosstalk
created this way,
we modified the experimental setup  by adding a periscope
in front of the polarizer.
Figure  \ref{fig:crstk_incl} shows the crosstalk for sample SS1 in the case of linear
polarization for inclination angles between 0\textdegree and 15\textdegree.
While there is a clear increase of crosstalk with $\phi$,
the range of acceptable viewing angles $\theta$ is reduced by only 5\textdegree.

\begin{figure}[t]
\centering
  \vspace*{-0.5cm}
  \hspace*{-0.7cm}
  \includegraphics[width=0.55\textwidth]{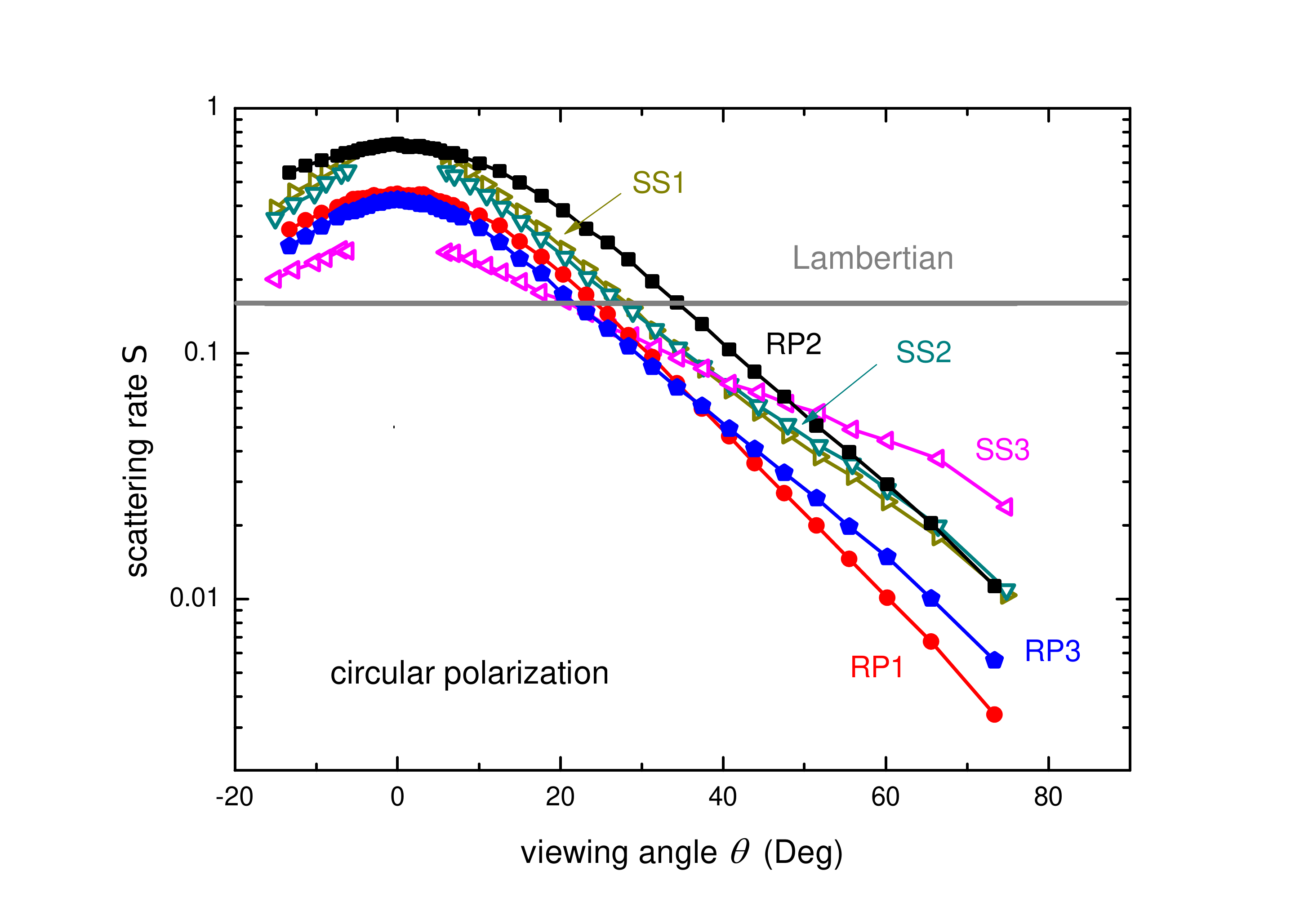}
  \vspace*{-0.9cm}
\caption{Angular dependence of the scattering rate measured with circular polarization.
Scattering rate values for linear polarization agree within 5.4 percent.}
\label{fig:sr}
\end{figure}

The scattering rates $S$ of the screen samples with circular polarization are shown in figure \ref{fig:sr}.
Deviations of $S$ measured with linear or the circular polarization are within our errorbars.
For high luminosities at small viewing angles SS1 and RP2 are the best choice, in terms of best homogeneity
SS3 comes closest to a Lambertian source.

\begin{figure}[b]
\begin{minipage}{17cm}
\centering
  \hspace*{0.5cm}
  \includegraphics[width=\textwidth]{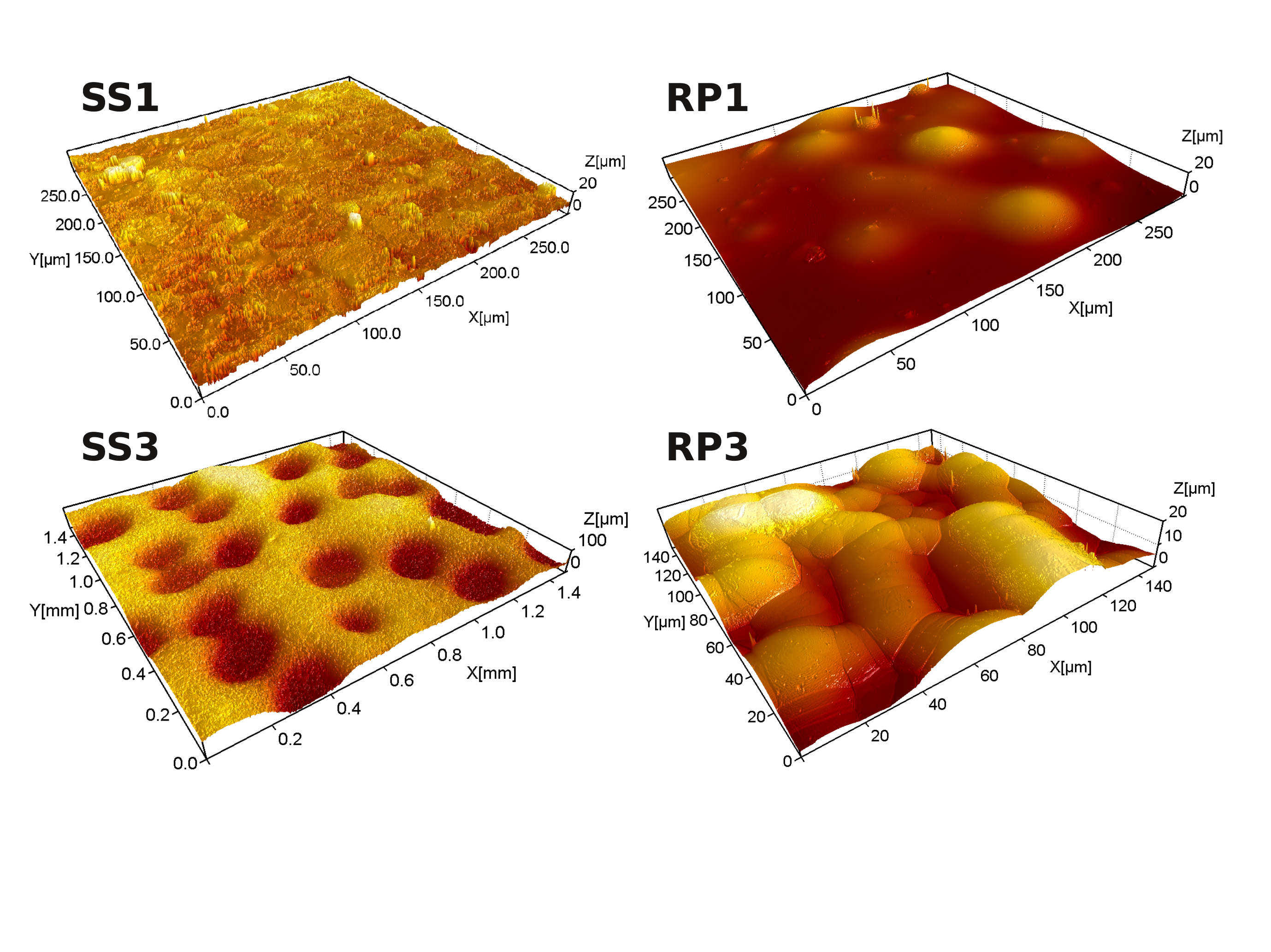}
    \vspace*{-2.7cm}
\caption{Perspective images of the surface texture of the screen samples.
Please note the different horizontal and vertical scales for sample SS3.
Images contain between 0.7\% (RP1) and 30\% (SS3) interpolated pixels.
}
\label{fig:roughness}
\end{minipage}
\end{figure}

From a theoretical side the performance of a screen will depend both on its material and its surface texture \cite{jin:10}.
While we do not have information on the electromagnetic properties of the screen material, the surface texture
can be measured with white light interferometry.
Perspective images of the surfaces of SS1, SS3, RP1, RP3  are shown in figure \ref{fig:roughness}.
The RMS (root mean square) roughness $R_q$ and the ratio between the surface area and the projected area $F$
of all six screen samples is listed in table \ref{tab:roughness}.

A comparison of the angular dependence of $S$ with these values hints at $R_q$ as a predictor for the deviation
from a Lambertian source. This is particularly shown by
SS3 which has by far the highest value of  $R_q$ and the smallest $\theta$ dependence of $S$.
Regarding the system crosstalk a similar correlation between  $R_q$ and the slope of $C$ at large angles might exist.
On the other side we do not find a clear correlation between the optical properties and $F$.

\newpage

\section{Conclusion}
All screens allow effective stereo projection for viewing angles up to 40\textdegree.
At larger angles the crosstalk of rear projection screens is considerably smaller than that of silver
screens. Also for each screen the crosstalk was larger with circular polarization than with linear.
However, when planing a display system additional factors have to be taken into account
like the available space behind the screen or the sensitivity of the system against the
viewers tilting their heads. Consequentially, no optimal solution for all possible scenarios exist.
While the roughness of the screens influences their large viewing angle behavior,
clearly more research is needed for a quantitative understanding.

{\it Acknowledgements:}
We would like to thank G\"unter Daszinnies from the company Screenlabs for providing the test samples.


\begin{thebibliography}{1}
\providecommand{\url}[1]{{#1}}
\providecommand{\urlprefix}{URL }
\expandafter\ifx\csname urlstyle\endcsname\relax
  \providecommand{\doi}[1]{DOI~\discretionary{}{}{}#1}\else
  \providecommand{\doi}{DOI~\discretionary{}{}{}\begingroup
  \urlstyle{rm}\Url}\fi

\bibitem{brennesholtz:08}
Brennesholtz, M.S., Stupp, E.H.: Projection Displays.
\newblock Wiley (2008)

\bibitem{darbha:10}
Darbha, G., Sch\"afer, T., Heberling, F., L\"uttge, A., Fischer, C.: Retention
  of latex colloids on calcite as a function of surface roughness and
  topography.
\newblock Langmuir \textbf{26}, 4743--4752 (2010)

\bibitem{huang:03}
Huang, K.C., Yuan, J.C., Tsai, C.H., Hsueh, W.J., Wang, N.Y.: A study of how
  crosstalk affects stereopsis in stereoscopic displays.
\newblock In: Proceedings of SPIE-IS\&T, vol. 5006, pp. 247--253 (2003)

\bibitem{iizuka:06}
Iizuka, K.: Welcome to the wonderful world of {3D:} introduction, principles
  and history.
\newblock Optics and Photonics News \textbf{17}, 42--51 (2006)

\bibitem{janssen:08}
Janssen, J.K.: {3D 2.0, Neuer Anlauf f\"ur Stereoskopie im Kino}.
\newblock {c't} \textbf{16}, 72--75 (2008)

\bibitem{jin:10}
Jin, L., Kasahara, M., Gelloz, B., Takizawa, K.: Polarization properties of
  scattered light from macrorough surfaces.
\newblock Optics Letters \textbf{35}, 595--597 (2010)

\bibitem{kim:10}
Kim, S.C., Kim, E.S.: Performance analysis of stereoscopic three-dimensional
  projection display systems.
\newblock {3D} Research \textbf{1}, 1--16 (2010)

\bibitem{richards:10}
Richards, M., Schnuelle, D.: The effective gain of a projection screen in an
  auditorium.
\newblock {SMPTE} Motion Imaging Journal \textbf{119}, 62 --67 (2010)

\bibitem{woods:11}
Woods, A.J.: How are crosstalk and ghosting defined in the stereoscopic
  literature?
\newblock In: Proceedings of SPIE-IS\&T, vol. 7863, p. 78630Z (2011)

\end{thebibliography}
\end{document}